\documentclass{article}
\usepackage{spconfa4,amsmath,graphicx,cite}
\usepackage{multirow}

\usepackage{amsfonts,amssymb,amsthm,algorithmic}
\usepackage{bm}
\usepackage{color}
\theoremstyle{definition}

\newtheorem*{lemma*}{Lemma}

\newtheorem{prop*}{Proposition*}

\newtheorem{definition*}{Definition*}

\newtheorem{note*}{Note*}

\newtheorem{theorem*}{Theorem*}

\newtheorem{col*}{Corollary*}

\title{FastFCA-AS: 
joint diagonalization based acceleration
of full-rank spatial covariance analysis for separating any number of sources
}
\name{Nobutaka Ito, Tomohiro Nakatani}
\address{NTT Communication Science Laboratories, NTT Corporation, Kyoto, Japan\\
\{ito.nobutaka,  nakatani.tomohiro\}@lab.ntt.co.jp}
\begin{document}
\maketitle
\ninept
\begin{abstract}
Here we propose {\it FastFCA-AS}, 
an accelerated algorithm for {\it Full-rank spatial Covariance Analysis (FCA)}, which is
a robust audio source separation method proposed by
Duong {\it et al.} [``Under-determined reverberant audio source
separation using a full-rank spatial covariance model,'' {\it IEEE Trans. ASLP}, vol.~18, no.~7, pp.~1830--1840, Sept. 2010].
In the conventional FCA, matrix inversion and matrix multiplication are required at each time-frequency point in each iteration of an iterative parameter estimation algorithm.
This causes a heavy computational load, thereby rendering the FCA infeasible in many applications. 
To overcome this drawback, we take
a joint diagonalization approach, whereby 
matrix inversion and matrix multiplication are reduced
to mere inversion and multiplication of 
diagonal entries.
This makes the FastFCA-AS significantly faster than the FCA
and even applicable to
observed data of long duration
or a situation with restricted computational resources. 
Although we have already proposed
another acceleration of the FCA for two sources,
the proposed FastFCA-AS is applicable to an arbitrary number of sources.
In an experiment with three sources and three microphones, the FastFCA-AS was over 
420 times faster than the FCA with a slightly better source separation performance.
\end{abstract}
\vspace{-2mm}\section{Introduction}
\label{sec:intro}

Duong {\it et al.}~\cite{Duong2010} have proposed a robust audio source separation method, which is called
{\it Full-rank spatial Covariance Analysis (FCA)} in this paper.
The FCA performs source separation by using the multichannel Wiener filter optimal in the Minimum Mean Square Error (MMSE) sense. To design the multichannel Wiener filter properly, it is crucial to accurately estimate the covariance matrices of the source signals.
In the FCA, these covariance matrices are estimated from the observed signals 
 by the maximum likelihood method based on the Expectation-Maximization (EM) algorithm.
A major drawback of the FCA is expensive computation.
Indeed, the above EM algorithm involves inversion and multiplication of covariance matrices at each time-frequency point in each iteration.
Since each of these matrix operations requires computation of complexity $O(I^3)$ ($I$: the matrix order) and the number of time-frequency points is normally huge,
 the FCA suffers from a heavy computational load. This may render the FCA inapplicable to
observed data of long duration
or a situation with restricted computational resources, such as
 hearing aids, distributed microphone arrays, online speech enhancement, {\it etc}.

In the two-source case, 
the above issue is addressed 
by a recently developed accelerated algorithm
for the FCA based on joint diagonalization 
by the generalized eigenvalue problem~\cite{FastFCAarXiv,Ito2018EUSIPCOFastFCA}. 
This method exploits 
the well-known property that,
for diagonal matrices, 
matrix inversion and matrix multiplication are reduced to mere
inversion and multiplication of diagonal entries. Owing to this property, the joint diagonalization 
reduces
the computational complexity of
matrix inversion and matrix multiplication
 from $O(I^3)$ to $O(I)$.
Consequently, 
the computation time of the FCA is curtailed significantly.
However, this method has a significant drawback of being only applicable to two sources.
Hence, we hereafter refer to this method as {\it FastFCA-TS} (Fast FCA for Two Sources).

To accelerate the FCA even when the number of sources exceeds two,
here we propose {\it FastFCA-AS} (Fast FCA for an Arbitrary number of Sources).
 Since joint diagonalization based on the generalized eigenvalue problem  is inapplicable to such a case, we introduce an alternative way of joint diagonalization. 
Specifically, joint diagonalization of the covariance matrices of the source signals is realized by 
maximum likelihood estimation of a basis-transform matrix for joint diagonalization 
and the diagonalized covariance matrices.
We propose a hybrid algorithm combining the EM algorithm and the fixed point iteration for the maximum likelihood parameter estimation.
Consequently, the proposed FastFCA-AS leads to significantly accelerated source separation
 even when the number of sources exceeds two.

We follow the following conventions in this paper. 
Signals are represented in the Short-Time Fourier Transform (STFT) domain, 
where the time and the frequency indices are denoted by $n$ and $f$ respectively. The number of frames is denoted by $N$, and the number of frequency bins up to the Nyquist frequency by $F$.
$\mathbf{0}$ denotes the column zero vector of an appropriate dimension,
$\mathbf{I}$ the identity matrix of an appropriate order,
$\text{diag}(\bm{\alpha})$ the diagonal matrix whose diagonal entries are given by the vector $\bm{\alpha}$,
$(\cdot)^\textsf{T}$ transposition,
$(\cdot)^\textsf{H}$ Hermitian transposition, 
$\text{tr}(\cdot)$ the trace, and 
$\det(\cdot)$ the determinant.
`$\alpha\triangleq \beta$' means that $\alpha$ is defined by $\beta$.

The rest of this paper is organized as follows. Section~\ref{sec:prob} formulates the source separation 
problem we deal with in this paper.
Section~\ref{sec:FCA} reviews the conventional FCA.
Section~\ref{sec:fastfca} describes the proposed FastFCA-AS.
Section~\ref{sec:exp} describes experimental evaluation, and
finally Section~\ref{sec:conc} concludes this paper.

\vspace{-2mm}\section{Problem Formulation}
\label{sec:prob}
Suppose $J$ source signals are observed by $I$ microphones.
Let $y_i(n,f)\in\mathbb{C}$ denote
the observed signal at the $i$th microphone and
$
\mathbf{y}(n,f)\triangleq\begin{bmatrix}
y_1(n,f)&y_2(n,f)&\dots&y_I(n,f)
\end{bmatrix}^\textsf{T}
$
the observed signals at all $I$ microphones.
We model $\mathbf{y}(n,f)$ by the sum of $J$ components $\mathbf{x}_j(n,f)\ (j=1,2,\dots,J)$ corresponding to the $J$ source signals:
$\mathbf{y}(n,f)=\sum_{j=1}^J\mathbf{x}_j(n,f)$. The components $\mathbf{x}_j(n,f)\ (j=1,2,\dots,J)$ are called {\it source images}.
The source separation problem we deal with in this paper is one of estimating  $\mathbf{x}_j(n,f)\ (j=1,2,\dots,J)$ from $\mathbf{y}(n,f)$. 

\vspace{-2mm}\section{FCA: Full-rank Spatial Covariance Analysis}
\label{sec:FCA}
\vspace{-2mm}\subsection{Full-Rank Spatial Covariance Model}

The FCA assumes that $\mathbf{x}_j(n,f)\ (j=1,2,\dots,J; n=1,2,\dots,N; f=1,2,\dots,F)$ independently follow the zero-mean complex Gaussian distribution:
\begin{equation}
p(\mathbf{x}_j(n,f))=\mathcal{N}(\mathbf{x}_j(n,f);\mathbf{0},\mathbf{R}_j(n,f)).\label{eq:srcmodel}
\end{equation}
Here, $\mathcal{N}(\bm{\alpha};\mathbf{m},\mathbf{R})$ denotes the complex Gaussian distribution with mean $\mathbf{m}$ and covariance matrix $\mathbf{R}$ for a random vector $\bm{\alpha}$, and
$\mathbf{R}_j(n,f)$
denotes the covariance matrix of $\mathbf{x}_j(n,f)$.
Importantly, $\mathbf{R}_j(n,f)$ is assumed to be parametrized as
\begin{align}
\mathbf{R}_j(n,f)=\underbrace{v_j(n,f)}_{\displaystyle \text{power spectrum}}\times\underbrace{\mathbf{S}_j(f)}_{\displaystyle \text{spatial characteristics}},\label{eq:fullrank}
\end{align}
where $\mathbf{S}_j(f)$ models the spatial characteristics of the $j$th source signal,
and $v_j(n,f)$ the power spectrum of the $j$th source signal.
The matrix
$\mathbf{S}_j(f)$ is called a {\it spatial covariance matrix}, and assumed to be Hermitian, positive definite (and thus full-rank).
The parameter $v_j(n,f)$ is assumed to be positive. 


\vspace{-2mm}\subsection{Maximum Likelihood Estimation of Model Parameters}
Once the model parameters $\mathbf{S}_j(f)$ and $v_j(n,f)$ have been obtained, the source image $\mathbf{x}_j(n,f)$ can be estimated, {\it e.g.,} by the MMSE estimator (also known as the multichannel Wiener filter):
\begin{equation}
\hat{\mathbf{x}}_j(n,f)= \mathbf{R}_j(n,f)\Biggl(\sum_{k=1}^J\mathbf{R}_k(n,f)\Biggr)^{-1}\mathbf{y}(n,f),\label{eq:MMSE0}
\end{equation}
where $\mathbf{R}_j(n,f)$
is given by (\ref{eq:fullrank}). 
Since the parameters $\mathbf{S}_j(f)$ and $v_j(n,f)$ are not known {\it a priori}, they are estimated from the observed signals by the maximum likelihood method. This amounts to  solving the following optimization problem:
\begin{align}
&\max_{\Theta}\ L_1(\Theta)\ \ \ \text{s.t.}\ \ \  \mathbf{S}_j(f)\succ 0, v_j(n,f)>0.\label{eq:opt}
\end{align}
Here, $\Theta$ denotes the ensemble of the parameters 
$\mathbf{S}_j(f)\ (j=1,2,\dots,J; f=1,2,\dots,F)$ and 
$v_j(n,f)\ (j=1,2,\dots,J;n=1,2,\dots,N;f=1,2,\dots,F)$
, and $L_1(\Theta)$ the log-likelihood function:
\begin{align}
L_1(\Theta)&\triangleq\sum_{n=1}^N\sum_{f=1}^F\ln\mathcal{N}\Biggl(\mathbf{y}(n,f);\mathbf{0},\sum_{j=1}^J v_j(n,f)\mathbf{S}_j(f)\Biggr).\label{eq:L1}
\end{align}
`$\mathbf{A}\succ 0$' means that $\mathbf{A}$ is a positive definite Hermitian matrix.

\vspace{-2mm}\subsection{Expectation-Maximization Algorithm}
The FCA realizes the maximum likelihood estimation by the EM algorithm~\cite{Dempster1977}, 
in which an Expectation step (E-step) and an Maximization step (M-step) are iterated alternately.

In the E-step, the current estimates of the parameters $\mathbf{S}_j(f)$ and $v_j(n,f)$ are used to update 
the posterior probability $p(\mathbf{x}_j(n,f)\mid \mathbf{y}(n,f))$ of
$\mathbf{x}_j(n,f)$, which turns out to be a complex Gaussian distribution again:
\begin{equation}
p(\mathbf{x}_j(n,f)\mid \mathbf{y}(n,f))=\mathcal{N}(\mathbf{x}_j(n,f);\bm{\mu}_j(n,f),\bm{\Phi}_j(n,f)),\label{eq:post}
\end{equation}
where $\bm{\mu}_j(n,f)$ denotes the mean and 
$\bm{\Phi}_j(n,f)$ the covariance matrix.
Therefore, the E-step amounts to updating $\bm{\mu}_j(n,f)$ and  $\bm{\Phi}_j(n,f)$. This is done by the following update rules:
\begin{align}
\bm{\mu}_j(n,f)&\leftarrow \mathbf{R}_j(n,f)\Biggl(\sum_{k=1}^J\mathbf{R}_k(n,f)\Biggr)^{-1}\mathbf{y}(n,f),\label{eq:MMSE}\\
\bm{\Phi}_j(n,f)&\leftarrow \mathbf{R}_j(n,f)-\mathbf{R}_j(n,f)
\Biggl(\sum_{k=1}^J\mathbf{R}_k(n,f)\Biggr)^{-1}\mathbf{R}_j(n,f),\label{eq:postquadexp}
\end{align}
where $\mathbf{R}_j(n,f)$
is given by (\ref{eq:fullrank}). 
Note that (\ref{eq:MMSE}) coincides with the MMSE estimator in (\ref{eq:MMSE0}).

In
the M-step, the estimates of the parameters $\mathbf{S}_j(f)$ and $v_j(n,f)$ are updated using $\bm{\mu}_j(n,f)$ and $\bm{\Phi}_j(n,f)$ obtained in the E-step. The update rules are as follows:
\begin{align}
v_j(n,f)&\leftarrow \frac{1}{I}\text{tr}\Bigl(\mathbf{S}_j(f)^{-1}(\bm{\mu}_j(n,f)\bm{\mu}_j(n,f)^\textsf{H}+\bm{\Phi}_j(n,f))\Bigr),\label{eq:vupdate}\\
\mathbf{S}_j(f)&\leftarrow \frac{1}{N}\sum_{n=1}^N\frac{1}{v_j(n,f)}(\bm{\mu}_j(n,f)\bm{\mu}_j(n,f)^\textsf{H}+\bm{\Phi}_j(n,f)).\label{eq:Supdate}
\end{align}

\vspace{-2mm}\subsection{Drawback}
\label{sec:drawback}
A major drawback of the FCA is expensive computation.
Indeed, each iteration of the above EM algorithm 
requires matrix inversion and matrix multiplication at each time-frequency point
as seen from (\ref{eq:MMSE}) and (\ref{eq:postquadexp}). 
Indeed, each iteration requires $(J+N)F$ matrix inversions and $2JNF$ matrix multiplications.
For example, for the experimental setting in
Section~\ref{sec:exp}: $I=J = 3$; $N = 249$; $F = 512$,
the number of matrix inversions is $(J + N)F = 129024$ per iteration, and
the number of matrix multiplications is $2JNF = 764928$ per iteration.


\vspace{-2mm}\section{FastFCA-AS: accelerated FCA for An Arbitrary Number of Sources}
\label{sec:fastfca}
\vspace{-2mm}\subsection{Approach: Joint Diagonalization}
This section describes the proposed FastFCA-AS, an accelerated version of the FCA applicable to an arbitrary number of sources.
The FastFCA-AS exploits the well-known fact that,
for diagonal matrices, 
matrix inversion and matrix multiplication are reduced to mere
inversion and multiplication of diagonal entries, which are both of complexity $O(I)$ instead of $O(I^3)$. This implies that,
if $\mathbf{R}_j(n,f)\,(j=1,2,\dots,J)$ were all diagonal, matrix inversion and matrix multiplication in (\ref{eq:MMSE}) and (\ref{eq:postquadexp}) would be reduced  to mere
inversion and multiplication of diagonal entries.
However, elements of
$\mathbf{x}_j(n,f)$ (that is, the $j$th source signal observed at different microphones)
are normally mutually correlated, which implies that its covariance matrix
$\mathbf{R}_j(n,f)$ has non-zero off-diagonal entries.

This motivates us to consider
joint diagonalization of the spatial covariance matrices $\mathbf{S}_j(f)\ (j=1,2,\dots,J)$.
That is, we consider transforming 
$\mathbf{S}_j(f)\ (j=1,2,\dots,J)$
into some diagonal matrices 
$\bm{\Lambda}_j(f)\ (j=1,2,\dots,J)$
by a single non-singular matrix $\mathbf{P}(f)$ 
as follows:
\begin{align}
\begin{cases}
\mathbf{P}(f)^\textsf{H}\mathbf{S}_1(f)\mathbf{P}(f)=\bm{\Lambda}_1(f),\\
\cdots\cdots\\
\mathbf{P}(f)^\textsf{H}\mathbf{S}_J(f)\mathbf{P}(f)=\bm{\Lambda}_J(f).
\end{cases}
\label{eq:diagcond}
\end{align}

For $J=2$ sources, the generalized eigenvalue problem yields $\mathbf{P}(f)$ and $\bm{\Lambda}_j(f)$ that satisfy (\ref{eq:diagcond})~\cite{Golub1983}. 
In the recently developed FastFCA-TS~\cite{FastFCAarXiv,Ito2018EUSIPCOFastFCA}, this approach is employed to accelerate the FCA without degrading the source separation performance.
However, the FastFCA-TS is limited to the two-source case. 

For more than two sources, the generalized eigenvalue problem based approach is inapplicable. Instead, in the proposed FastFCA-AS,
$\mathbf{S}_j(f)$ is assumed to be parametrized as
\begin{align}
\begin{cases}
\mathbf{S}_1(f)=(\mathbf{P}(f)^{-1})^\textsf{H}\bm{\Lambda}_1(f)\mathbf{P}(f)^{-1},\\
\cdots\cdots\\
\mathbf{S}_J(f)=(\mathbf{P}(f)^{-1})^\textsf{H}\bm{\Lambda}_J(f)\mathbf{P}(f)^{-1}.
\end{cases}
\label{eq:diagconst}
\end{align}
(\ref{eq:diagconst}) is obtained by solving (\ref{eq:diagcond}) for $\mathbf{S}_j(f)$. 
The parameters $\mathbf{P}(f)$, $\bm{\Lambda}_j(f)$, and $v_j(n,f)$ are estimated from the observed signals by the maximum likelihood method.
This makes it possible to accelerate the FCA even for more than two sources.

\vspace{-2mm}\subsection{Objective Function}
The maximum likelihood method amounts to solving the following optimization problem:
\begin{align}
&\max_{\Psi}\ L_2(\Psi)\ \ \ \notag\\
&\ \text{s.t.}\ \ \ \mathbf{P}(f)\in \text{GL}(I,\mathbb{C}), \bm{\Lambda}_j(f)\succ 0\text{: diagonal}, v_j(n,f)>0.\hspace{-2mm}
\end{align}
Here, $\Psi$ denotes the ensemble of the parameters
$\mathbf{P}(f)\ (f=1,2,\dots,F)$,
$\bm{\Lambda}_j(f)\ (j=1,2,\dots,J;f=1,2,\dots,F)$,
and $v_j(n,f)\ (j=1,2,\dots,J;n=1,2,\dots,N;f=1,2,\dots,F)$.
$L_2(\Psi)$ denotes the log-likelihood function:
\begin{align}
&L_2(\Psi)\triangleq\sum_{n=1}^N\sum_{f=1}^F\ln\mathcal{N}\Biggl(\mathbf{y}(n,f);\mathbf{0},\notag\\
&\phantom{L_2(\Psi)\triangleq\sum_{n=1}^N\sum_{f=1}^F\ln\mathcal{N}\Biggl(}
\sum_{j=1}^J v_j(n,f)(\mathbf{P}(f)^{-1})^\textsf{H}\bm{\Lambda}_j(f)\mathbf{P}(f)^{-1}\Biggr).\label{eq:L2}
\end{align}
$\text{GL}(I,\mathbb{C})$ denotes
the set of  the non-singular complex matrices of order $I$.

\vspace{-2mm}\subsection{Optimization Algorithm}
The FastFCA-AS realizes the maximum likelihood estimation 
by a hybrid algorithm combining the EM algorithm and the fixed point iteration.
In this algorithm, the following two steps are alternated:
\begin{enumerate}
\item Update $\bm{\Lambda}_j(f)\ (j=1,2,\dots,J; f=1,2,\dots,F)$ and $v_j(n,f)\ (j=1,2,\dots,J;n=1,2,\dots,N;f=1,2,\dots,F)$ by applying one iteration of the EM algorithm.
\item Update $\mathbf{P}(f)\ (f=1,2,\dots,F)$ by the fixed point iteration.
\end{enumerate}


\subsubsection{EM-Based $\bm{\Lambda}_j(f)$ and $v_j(n,f)$ Update}
\label{sec:EM}
The EM-based $\bm{\Lambda}_j(f)$ and $v_j(n,f)$ update consists of the E-step and the M-step described in the following.

In the E-step, the posterior probability $p(\mathbf{x}_j(n,f)\mid \mathbf{y}(n,f))$ of $\mathbf{x}_j(n,f)\ (j=1,2,\dots,J; n=1,2,\dots,N;f=1,2,\dots,F)$ is updated based on the current parameter estimates.
As in the conventional FCA,
$p(\mathbf{x}_j(n,f)\mid \mathbf{y}(n,f))$ turns out to be a complex Gaussian distribution given by (\ref{eq:post})
with 
the mean $\bm{\mu}_j(n,f)$ given by (\ref{eq:MMSE})
and the covariance matrix $\bm{\Phi}_j(n,f)$ by (\ref{eq:postquadexp}).
Unlike the FCA, however,
$\mathbf{R}_j(n,f)$ in (\ref{eq:MMSE})
and (\ref{eq:postquadexp}) is given by
\begin{equation}
\mathbf{R}_j(n,f)=v_j(n,f)
(\mathbf{P}(f)^{-1})^\textsf{H}\bm{\Lambda}_j(f)\mathbf{P}(f)^{-1}.\label{eq:covmodelprop}
\end{equation}
Substitution of (\ref{eq:covmodelprop}) into (\ref{eq:MMSE}) and (\ref{eq:postquadexp}) 
yields
\begin{align}
&\underbrace{\mathbf{P}(f)^\textsf{H}\bm{\mu}_j(n,f)}_{\displaystyle \tilde{\bm{\mu}}_j(n,f)}\notag\\
&=v_j(n,f)\bm{\Lambda}_j(f)\Biggl(\sum_{k=1}^Jv_k(n,f){\bm{\Lambda}}_k(f)\Biggr)^{-1}
\underbrace{\mathbf{P}(f)^\textsf{H}\mathbf{y}(n,f)}_{\displaystyle \tilde{\mathbf{y}}(n,f)},\label{eq:mutilde0}\\
&\underbrace{\mathbf{P}(f)^\textsf{H}{\mathbf{\Phi}}_j(n,f)\mathbf{P}(f)}_{\displaystyle \tilde{\bm{\Phi}}_j(n,f)}= v_j(n,f)\bm{\Lambda}_j(f)\notag\\
&-v_j(n,f)\bm{\Lambda}_j(f)
\Biggl(\sum_{k=1}^Jv_k(n,f)\bm{\Lambda}_k(f)\Biggr)^{-1}(v_j(n,f)\bm{\Lambda}_j(f)).\label{eq:Phitilde0}
\end{align}
Therefore, $\tilde{\bm{\mu}}_j(n,f)$ and 
$\tilde{\bm{\Phi}}_j(n,f)$,
basis-transformed versions of $\bm{\mu}_j(n,f)$ and $\bm{\Phi}_j(n,f)$,
can be updated 
by (\ref{eq:mutilde0}) and (\ref{eq:Phitilde0}), 
in which matrix inversion and matrix multiplication are of complexity $O(I)$ instead of $O(I^3)$ 
owing to the joint diagonalization.

In the M-step, $v_j(n,f)\ (j=1,2,\dots,J; n=1,2,\dots,N; f=1,2,\dots,F)$ and $\mathbf{\Lambda}_j(f)\ (j=1,2,\dots,J; f=1,2,\dots,F)$ are updated based on
maximization of the following Q-function:
\begin{align}
Q(\Psi)
&=-\sum_{n=1}^N\sum_{f=1}^F\sum_{j=1}^J
\Biggl[
\ln\det \bigl(v_j(n,f)(\mathbf{P}(f)^{-1})^\textsf{H}\bm{\Lambda}_j(f)\mathbf{P}(f)^{-1}\bigr)\notag\\
&\phantom{=}+\text{tr}\Bigl(
(v_j(n,f)\bm{\Lambda}_j(f))^{-1}\bigl(\tilde{\bm{\Phi}}_j(n,f)+
\tilde{\bm{\mu}}_j(n,f)\tilde{\bm{\mu}}_j(n,f)^\textsf{H}\bigr)
\Bigr)
\Biggr].
\end{align}
Partial differentiation with respect to $v_j(n,f)$ and $\bm{\Lambda}_j(f)$ leads to the following update rules:
\begin{align}
v_j(n,f)&\leftarrow\frac{1}{I}\text{tr}\Bigl(\bm{\Lambda}_j(f)^{-1}(\text{diag}(|\tilde{\bm{\mu}}_j(n,f)|^2)+\tilde{\bm{\Phi}}_j(n,f))\Bigr),\label{eq:updatev}\\
\bm{\Lambda}_j(f)&\leftarrow\frac{1}{N}\sum_{n=1}^N\frac{1}{v_j(n,f)}(\text{diag}(|\tilde{\bm{\mu}}_j(n,f)|^2)+\tilde{\bm{\Phi}}_j(n,f)),\label{eq:Lambdaupdate}
\end{align}
where $|\cdot|^2$ is computed in an entry-wise manner.

\subsubsection{Fixed Point Iteration Based $\mathbf{P}(f)$ Update}
\label{sec:FPI}
The basis-transform matrix $\mathbf{P}(f)\ (f=1,2,\dots,F)$
is updated based on the fixed point iteration 
applied to the log-likelihood function (\ref{eq:L2}). Partial differentiation (the matrix Wirtinger derivative~\cite{Hjorungnes2007})
of (\ref{eq:L2}) with respect to the complex conjugate $\mathbf{P}(f)^\ast$ of $\mathbf{P}(f)$ is given by 
\begin{align}
\frac{\partial L_2(\Psi)}{\partial\mathbf{P}(f)^\ast}&=N(\mathbf{P}(f)^{-1})^\textsf{H}\notag\\
&\phantom{=}-
\sum_{n=1}^N\mathbf{y}(n,f)\mathbf{y}(n,f)^\textsf{H}\mathbf{P}(f)\Biggl(\sum_{j=1}^Jv_j(n,f)\bm{\Lambda}_j(f)\Biggr)^{-1}.\label{eq:partialP}
\end{align}
Setting (\ref{eq:partialP}) to zero and vectorizing both sides of the equation yields
\begin{align}
\text{vec}(\mathbf{P}(f))
&=\Biggl[\frac{1}{N}\sum_{n=1}^N
\Biggl(\sum_{j=1}^Jv_j(n,f)
\bm{\Lambda}_j(f)\Biggr)^{-1}\notag\\
&\phantom{=\Biggl[}\otimes\bigl(\mathbf{y}(n,f)\mathbf{y}(n,f)^\textsf{H}\bigr)\Biggr]^{-1}\text{vec}\bigl((\mathbf{P}(f)^{-1})^\textsf{H}\bigr)\label{eq:vecP}
\end{align}
owing to the formula $\text{vec}(\mathbf{AXB})=(\mathbf{B}^\textsf{T}\otimes
\mathbf{A})\text{vec}(\mathbf{X})$.
Here, $\text{vec}$ denotes the operator that stacks
the column vectors of the input matrix, and $\otimes$ the Kronecker product.
Noting the block diagonal structure, we can rewrite (\ref{eq:vecP}) as follows:
\begin{align}
[\mathbf{P}(f)]_i&\leftarrow \Biggl[
\frac{1}{N}\sum_{n=1}^N\frac{1}{\sum_{j=1}^Jv_j(n,f)[\bm{\Lambda}_j(f)]_{ii}}\mathbf{y}(n,f)\mathbf{y}(n,f)^\textsf{H}
\Biggr]^{-1}\notag\\
&\phantom{\leftarrow}\times\bigl[(\mathbf{P}(f)^{-1})^\textsf{H}\bigr]_i.\label{eq:FPI}
\end{align}
Here, $[\mathbf{A}]_i$ denotes the $i$th column of the matrix $\mathbf{A}$,
and $[\mathbf{A}]_{il}$ the $(i,l)$-entry of the matrix $\mathbf{A}$.
The fixed point iteration
consists in iterating (\ref{eq:FPI}).

\vspace{-2mm}\subsection{Advantage}
The proposed FastFCA includes only $(I+1)FK$ matrix inversions per iteration of the hybrid algorithm and no matrix multiplications, 
where $K$ denotes the number of iterations in the fixed point iteration.
Note that, unlike the FCA, the number of matrix inversions does not
depend on $N$, which is typically large.
Here, matrix inversions and matrix multiplications for diagonal matrices were not counted, because their computational complexity is $O(I)$ instead of $O(I^3)$.
For the experimental setting in
Section~\ref{sec:exp} where $K=1$, the number of matrix inversions is only
$(I+1)FK=2048$ per iteration of the hybrid algorithm.

\vspace{-2mm}\subsection{Discussion}
Here we described the hybrid algorithm combining the EM algorithm and the fixed point iteration.
Other optimization techniques could also be employed. For example, the fixed point iteration
for updating $\mathbf{P}(f)$ could be replaced by
 the gradient method, the natural gradient method, Newton's method, {\it etc.}
We could also employ the normal EM algorithm,
in which $\mathbf{P}(f)$ is also updated in the M-step.

\begin{figure}[tb]\centering
\includegraphics[width=0.8\columnwidth]{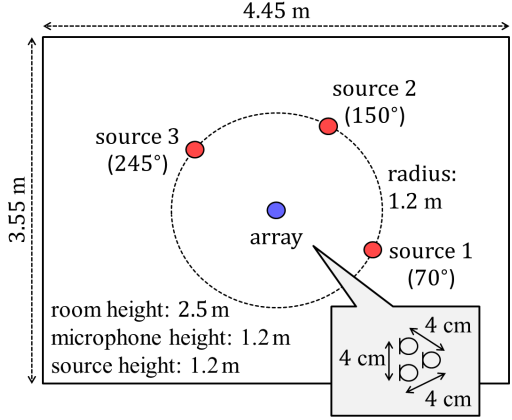}\vspace{-2mm}
\caption{Experimental setting (bird's eye view).}
\label{fig:setting}
\end{figure}

\begin{table}
\centering
\caption{Experimental conditions.}
\label{tab:cond}
\begin{tabular}{ll}\hline
sampling frequency &16\,kHz\\
frame length&1024 (64\,ms)\\
frame shift&512 (32\,ms)\\
window&square root of Hann\\
number of iterations&20\\
\hline
\end{tabular}
\end{table}

\begin{figure}[t]
\centering
\includegraphics[width=\columnwidth]{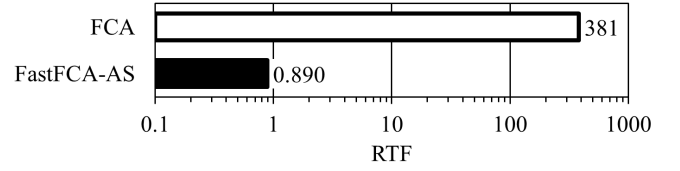}\vspace{-4mm}
\caption{Real Time Factor (RTF).}\vspace{-3mm}
\label{fig:RTF}
\end{figure}

\begin{figure}[tb]
\centering\includegraphics[width=\columnwidth]{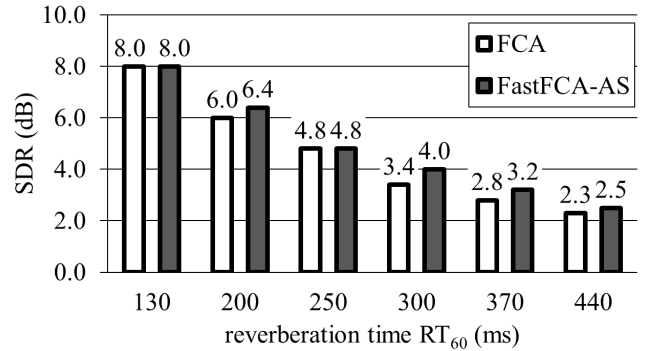}\vspace{-2mm}
\caption{Signal-to-Distortion Ratio (SDR).}
\label{fig:SDR}
\end{figure}

\vspace{-2mm}\section{Experimental Evaluation}
\label{sec:exp}
We conducted a source separation experiment to compare
the proposed FastFCA-AS with the FCA~\cite{Duong2010} (see Section~\ref{sec:FCA}).
These methods were implemented in MATLAB (R2013a) and run on an Intel i7-2600 3.4-GHz octal-core CPU.
Observed signals were generated by convolving 8\,s-long
English speech signals with room impulse responses~\cite{Sawada2011TASLP}
measured in an experiment room. 
The locations of the sources and the microphones are 
depicted in Fig.~\ref{fig:setting}.
The reverberation time $\text{RT}_{60}$ was 130, 200, 250, 300, 370, or 440\,ms, and for each reverberation time,
ten trials were conducted with different combinations of speech signals.
The parameters were initialized based on mask-based covariance matrix estimation~\cite{MehrezMMSE,Yoshioka2015} with the masks obtained by the method in \cite{Sawada2011TASLP}.
The source images were estimated using the multichannel Wiener filter in all algorithms.
Some other conditions are found in
Table~\ref{tab:cond}.

Figure~\ref{fig:RTF} shows the Real Time Factor (RTF) of the parameter estimation  averaged over all ten trials and all six reverberation times, and
Figure~\ref{fig:SDR} shows the Signal-to-Distortion Ratio (SDR)~\cite{Vincent2006} averaged over all three sources and all ten trials.
The proposed FastFCA-AS was over 420 times faster than the FCA
with its source separation performance slightly better than the FCA.

\vspace{-2mm}\section{Conclusions}
In this paper, we have proposed the FastFCA-AS, an accelerated algorithm for the FCA.
Compared to the conventional FastFCA-TS, 
the FastFCA-AS has a major advantage 
of being applicable to not only two sources but also more than two sources.

\label{sec:conc}

\newpage
\bibliographystyle{IEEEbib}

\end{document}